\pdfoutput=1
\documentclass{article}

\usepackage[english]{babel}
\usepackage[utf8]{inputenc}
\usepackage{color}

\usepackage{graphicx} 
\usepackage{amsmath} 
\usepackage{amssymb} 
\usepackage{amsthm} 
\usepackage{tabularx} 
\usepackage{siunitx} 
\usepackage{algorithm} 
\usepackage[noend]{algpseudocode} 
\usepackage{caption} 
\usepackage{subcaption} 
\usepackage{filecontents} 

\usepackage{authblk}

\everymath{\displaystyle}
\graphicspath{{img/}}

\title{ Statistical behaviour of interfaces subjected to curvature flow and torque effects applied to microstructural evolutions. }
\author[1]{Sebastian~Florez}
\author[1]{Karen~Alvarado}
\author[1]{Brayan~Murgas}
\author[1]{Nathalie~Bozzolo}
\author[2]{Dominique~Chatain}
\author[3]{Carl E.~Krill III}
\author[3]{Mingyan~Wang}
\author[4]{Greg S.~Rhorer}
\author[1]{Marc~Bernacki\thanks{corresponding author}}

\affil[1]{Mines-ParisTech, PSL-Research University, CEMEF – Centre de mise en forme des mat\'{e}riaux, CNRS UMR 7635, CS 10207 rue Claude Daunesse, 06904 Sophia Antipolis Cedex, France}%
\affil[2]{Aix-Marseille Univ, CNRS, CINaM, 13009 Marseille, France}%
\affil[3]{Institute of Functional Nanosystems, Ulm University, 89081 Ulm, Germany}%
\affil[4]{Department of Materials Science and Engineering, Carnegie Mellon University, Pittsburgh, Pennsylvania 15213, USA}%

\begin{document}
\maketitle

\section*{Abstract}

The movement of grain boundaries in pure metals and alloys with a low concentration of dislocations has been historically proved to follow \emph{curvature flow} behavior. This mechanism is typically known as grain growth (GG). However, recent 3D in-situ experimental results tend to question this global picture concerning the influence of the curvature on the kinetics of interface migration. 
This article explains, thanks to 2D anisotropic full-field simulations, how the torque effects can complexify these discussions.
 It is then illustrated that neglecting torque effects in full-field formulations remains potentially a strong hypothesis. The apparent reduced mobility can be much more complex than expected without necessarily questioning the influence of the curvature on the local kinetic equation.

\section{Introduction}\label{sec:introduction}
Grain boundaries (GB) are ubiquitous components in functional and structural materials which influence their electrical, magnetic, thermal, or mechanical properties. Mastering a GB network and its evolution during thermomechanical treatment is an absolute requirement for optimizing materials in most technological applications. Three main processes lead to the global minimization of the system energy during thermomechanical treatments of monophase polycrystals: recovery, recrystallization, and grain growth (GG). Although the GG mechanism has been known for more than one century and actively studied for 70 years \cite{Smith1948, Burke1952, Mullins1956, vonNeumann1952, Herring1951}, it remains a very active research topic in terms of involved equations, and modeling \cite{rollett2017recrystallization}. This historical background has brought us to the current global picture: GG in single-phase polycrystals relatively free of point and line defects is driven by the minimisation of GB areas to decrease the system's interfacial energy. The velocity $\vec{v}$ at which GBs move when subjected to a driving pressure $P$ is assumed to be the product of the GB mobility $\mu$ and $P$. This pressure is the product of the GB energy $\gamma$ and the GB curvature, i.e., the trace of the curvature tensor in the 3D space, $\kappa$. Thus, at the polycrystal scale, GG is classically described by the equation 
\begin{equation}\label{eq:basickinetic}
\vec{v}=-\mu\gamma\kappa\vec{n},
\end{equation}
where $\vec{n}$ is the outside normal to the GB and GB properties should be depicted in the 5-dimensional parameter space of the misorientation of the abutting crystals and the GB plane (inclination). Literature classically uses the term "reduced mobility" to describe the $\mu\gamma$ product. However, this picture remains questionable in many aspects when the local description of GB migration is aimed. The main question probably lies in the appropriateness of representing the migration of an interface using its macroscopic properties (misorientation and inclination). Thanks to the improvement of GG experiments, this aspect is more and more discussed in the state of the art. Indeed, monitored in-situ in a synchrotron by 3D X-ray diffraction microscopy techniques, which avoid bias inherent to 2D observations, opens the way to precise reverse engineering \cite{mingyan,Rohrer2021,ZHANG2020211} in order to discuss the $v/\kappa$ ratio, i.e., the apparent reduced mobility of each GB while knowing their characteristics in the 5-dimensional parameter space of the misorientation and inclination.
If Eq.\ref{eq:basickinetic} is indeed a simplification of lower scale phenomena in constant discussions \cite{PhysRevLett.119.246101,zhu2019situ}, the interpretation of the results obtained tends in some cases to refute or at least to question the real impact of the curvature on the kinetics of the interfaces or to dispel the 5-parameter description of the reduced mobility. Interfaces not moving towards their center of curvature ($v/\kappa$ ratio of opposite sign to the expected one) and/or presenting kinetics very different from other grain boundaries with similar characteristics in the 5-dimensional space can very legitimately raise the discussion.
The numerical results presented here illustrate how topological and torque effects can complicate grain boundary dynamics. If this aspect was already discussed using atomistic simulations \cite{ABDELJAWAD2018440, MOORE2021117220}, the aim here is to show that the ratio $v/\kappa$ per interface can present complex statistics at the polycrystal scale while respecting a curvature-flow type kinetic equation and 5-parameters reduced mobility. From a simulation results point of view, this is not trivial either. Indeed if polycrystal GG has been widely studied using a variety of numerical approaches like phase-field \cite{Miyoshi2017, moelans2009comparative}, Monte Carlo or Cellular Automata \cite{upmanyu2002boundary}, vertex \cite{BarralesMora2010}, FE-LS methods \cite{Bernacki2011, miessen2015advanced, Fausty2020, Hallberg2019}, front-tracking in a FE context \cite{Florez2020b,Florez2020d}; there is, however, a real confusion on the notion of anisotropic full-field simulation whereas these models can be used as fitting methodologies \cite{ZHANG2020211} of experimental 3D datasets. 
Indeed, it must be highlighted that the distinction between 3-parameter and 5-parameter space for the GB description in full-field frameworks is not straightforward. In the literature, heterogeneous values of GB properties are classically referred to as anisotropic. Moreover, in the rare formulations where both the effect of the misorientation and the inclination are taken into account, the GB energy dependence on the normal direction is generally defined without inquiring about the need for the torque terms in solving equations. Thus, the conclusions of many full-field simulation works published today are probably biased and contradictory either because of the complexity of the considered microstructures or because of the limitations of the simulations used to reproduce actual 5-parameter GB kinetics with torque effects. This article will illustrate that anisotropic simulations with or without taking into account torque terms can deviate substantially with regards to grain morphologies and local and global grain growth kinetics and $v/\kappa$ ratio.

\section{Numerical Method}\label{sec:numericalmethod_6}

In this section, some of the requirements to perform full-anisotropic computations for polycrystals are given. The so-called "Topological Remeshing in lAgrangian framework for large interface
Motion" (TRM) model presented in \cite{Florez2020b} and adapted to heterogeneous/anisotropic simulations in \cite{Florez2021a} will be used in this work. The TRM model allows the representation of a polycrystal with a 2D body-fitted unstructured mesh, and uses a Lagrangian scheme for the modeling of GB migration. The TRM model has allowed the FE study of different microstructural mechanisms such as grain growth (GG) \cite{Florez2020b} and dynamic recrystallization (DRX) \cite{Florez2020d}. The TRM model also features a well balanced parallel implementation that allows its use in an HPC environment \cite{Florez2020c}, with a gain in 2D computational performance 15 times faster than classical methods such as the LS-FE approach.\\

GB energy is classically considered as dependant of the tuple $(M_{lw}, \vec{n})$ where $M_{lw}$ represents the misorientation tensor of the two adjacent grains ($l$ and $w$) defining the GB and $\vec{n}$ is the local normal to the GB \cite{rollett2017recrystallization} pointed in the direction of the curvature center. As thus, $M_{lw}$ being constant for a given GB, while $\vec{n}$ taking different directions in a single GB, two kinds of anisotropic simulations can be defined \cite{Fausty2020, Fausty2020a,Hallberg2019}: the first, the \emph{heterogeneous} case, where only the dependence on $M_{lw}$ is taken into account (thus each GB takes a constant value of $\gamma$ disregarding its inclination), and the second the \emph{anisotropic} case where the dependence of $\gamma$ on the tuple $(M_{lw}, \vec{n})$ is considered (thus the $\gamma$ field can take different values along a non-flat GB).\\ 

The dependence on the misorientation matrix can be translated into a dependence on the disorientation $\theta$ and the misorientation axis attributed to each GB. The dependence to the misorientation axis will not be discussed here. The computation of $\theta$ depends on the alloy's crystallographic properties and is, in general, computed using a brute-force algorithm (see \cite{Florez2021a} for information regarding its computation in the context of the TRM model). Additionally, in this article, we will compute the inclination dependence by using the counter-clockwise angle $\omega$ measured between the projection on the XY-plane of the misorientation axis and the GB normal $\vec{n}$. As such, we define the value of $\gamma$ for each interface as follows:

\begin{equation}
\label{Equation_Gamma_Function}
\centering
\gamma_{(\theta, \omega)} =  \gamma^{D}_{(\theta)} \cdot \gamma^{I}_{(\omega)}, 
\end{equation}

where the terms $\gamma^{D}_{(\theta)}$ and $\gamma^{I}_{(\omega)}$ are the contribution to the energy given by the disorientation and by the inclination angle respectively. Typically, the value of $\gamma^{D}_{(\theta)}$ can be computed using a Read-Shockley formulation \cite{Read1950} or similar (see \cite{Fausty2020, Florez2021a} for comparisons of different $\gamma^{D}_{(\theta)}$ functions). Note that, in the following, the heterogeneous formulation will use $\gamma^{I}_{(\omega)}=1$.\\

In the absence of stored energy and with isotropic properties, grain boundary kinetics have historically been approximated by curvature flow, where the local curvature drives the local velocity of GBs as detailed in Eq.\ref{eq:basickinetic}. In an anisotropic context, however, the local velocity of an interface, driven by the minimization of the \emph{interfacial} energy, depends on the second derivative of $gamma$ as a function of the inclination angle ($w$), which allows writing \cite{BarralesMora2008}:
 
 \begin{equation}
\label{Eq:Equation_Gamma_Function}
\centering
\vec{v}=-\mu\kappa\vec{n}(\gamma+\frac{\partial\gamma^2}{\partial\omega^2}), 
\end{equation}

which is an approximation in 2D of the more general form of Herring's equations for single interfaces \cite{herring1999surface}. The sum of $\gamma$ and its second derivative with respect to inclination is also well-known as the grain boundary stiffness \cite{FOILES20063351,ABDELJAWAD2018440}.

This equation can be approximated for discrete interfaces using the following expression \cite{Florez2021a}: 

\begin{equation}
\label{Eq:Model2KawaVelocity}
\centering
\vec{v}_i=\mu_i\left({\frac{\sum_j^p\gamma_{ij} \vec{t}_{ij} + \tau_{ij}\vec{n}_{ij}}{\sum_j^p|\overline{N_iN_j}|\cdot p^{-1}}}\right),
\end{equation}

which is a variant of a discrete formulation proposed in \cite{BarralesMora2008} and where the index $i$ and $j$ are the identities of adjacent nodes $N_i$ and $N_j$ representing one of the segments $\overline{N_iN_j}$ of a given GB, $\mu_i$ is the mobility of node $N_i$ and which will be considered as homogeneous in the following ($\mu_i=\mu$), $\gamma_{ij}$, $t_{ij}$ and $\vec{n}_{ij}$ are respectively the boundary energy, the unit tangent vector and the normal of the segment $\overline{N_iN_j}$ and where $p$ is the number of segments connected to node $N_i$ (Eq.\ref{Eq:Model2KawaVelocity} also works for multiple junctions). Note the appearing of the term $\tau_{ij}$ which corresponds to the torque experienced by the segment $\overline{N_iN_j}$ due to the dependence on the inclination angle $\omega$ \cite{DeWit1959, herring1999surface}, this torque is defined as follows:

\begin{equation}
\label{Eq:TorqueTerm}
\centering
\tau_{ij}=-\frac{d\gamma}{d\omega_{ij}}.
\end{equation}

\section{Numerical Results}\label{sec:numericalresults_6}

\begin{figure}[!h]
\centering
\includegraphics[width=1.0\textwidth] {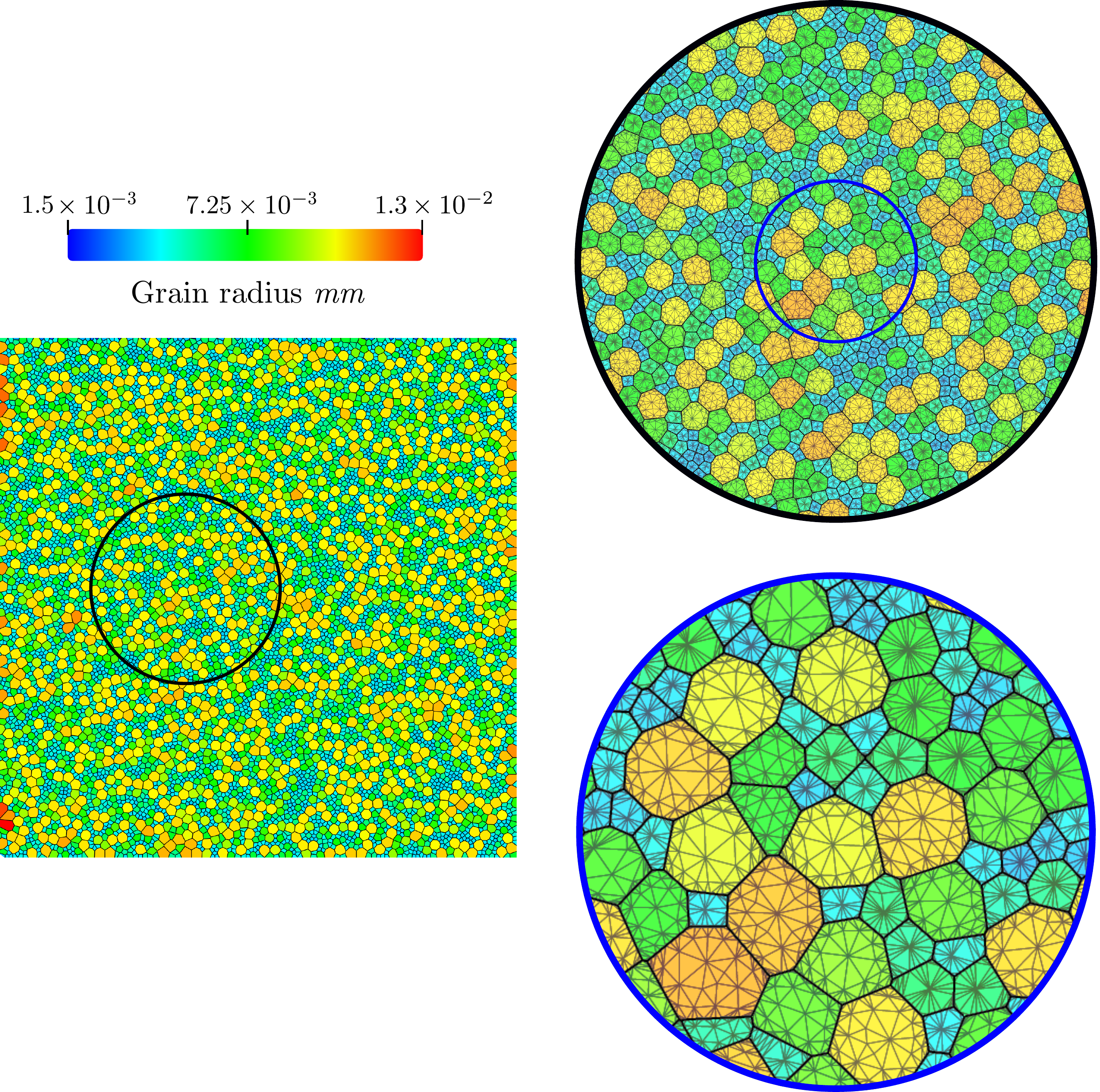}
\caption{Example of one of the four Laguerre-Voronoi tessellations used as initial states. Colors are representative of the individual grain radius (defined as $\sqrt{S/\pi}$ with $S$ the grain surface): (left) complete domain, (right) successive zooms and mesh used.}
\label{Figure_Example_InitialTess}
\end{figure}

\begin{figure}[!h]
\centering
\includegraphics[width=1.0\textwidth] {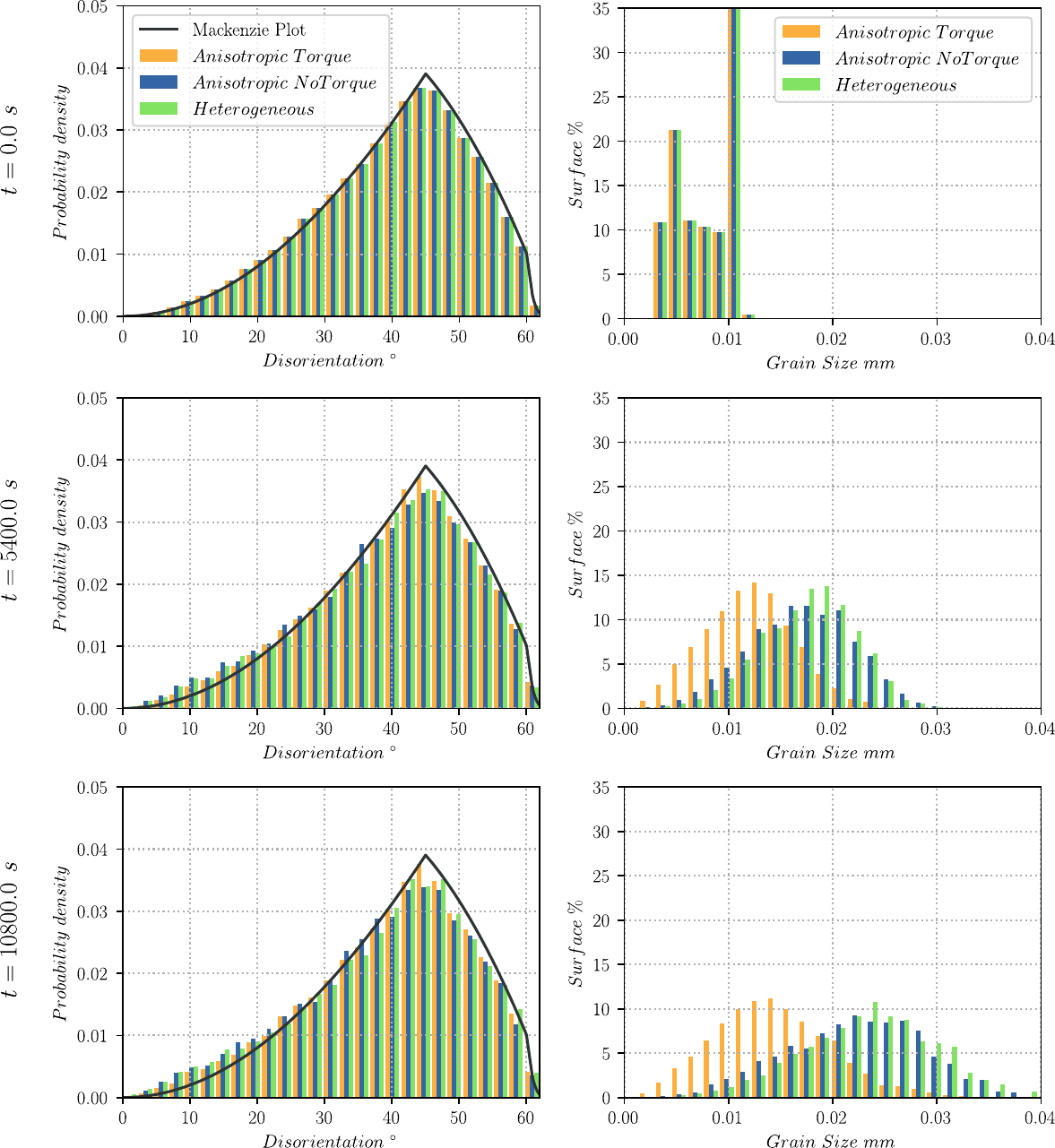}
\caption{Results of the test cases with different attributes for the computation of the GB velocities. Each case uses 4 tests with 4 different initial tessellations. (left) Normalized Mackenzie grain boundary disorientation plot \cite{Mackenzie1957} is given for reference. (Right) Grain size distributions as a percent of the total domain surface.}
\label{Figure_Mackenzie_Distribs_MeanSize}
\end{figure}

This section presents the results obtained by using the anisotropic TRM model detailed in \cite{Florez2021a}. Three sets of simulations were performed, the first using a heterogeneous configuration ($\gamma_{(\theta)}$), the second using an anisotropic configuration without consideration of torque terms (disregarding the second term in Eq.\ref{Eq:Equation_Gamma_Function}) but by considering the dependence of the grain boundary energy ($\gamma_{(\theta, \omega)}$) on the inclination angle, and the last one  using the same anisotropic configuration but considering torque terms. Each set contains four simulations using different initial states obtained from optimized Laguerre-Voronoi tessellations \cite{Hitti2012,Hitti2013}, generated with different random components but representing the same statistical distribution. Each representative volume element (RVE) corresponds to a square of $1\times 1~mm$ with about 8000 initial grains (see Fig.\ref{Figure_Example_InitialTess} (left)). The orientation of each grain was attributed at random, which typically results in a disorientation field following a Mackenzie probability distribution \cite{Mackenzie1957} (see Fig.\ref{Figure_Mackenzie_Distribs_MeanSize} (top-left)).\\

\begin{figure}[!h]
\centering
\includegraphics[width=1.0\textwidth] {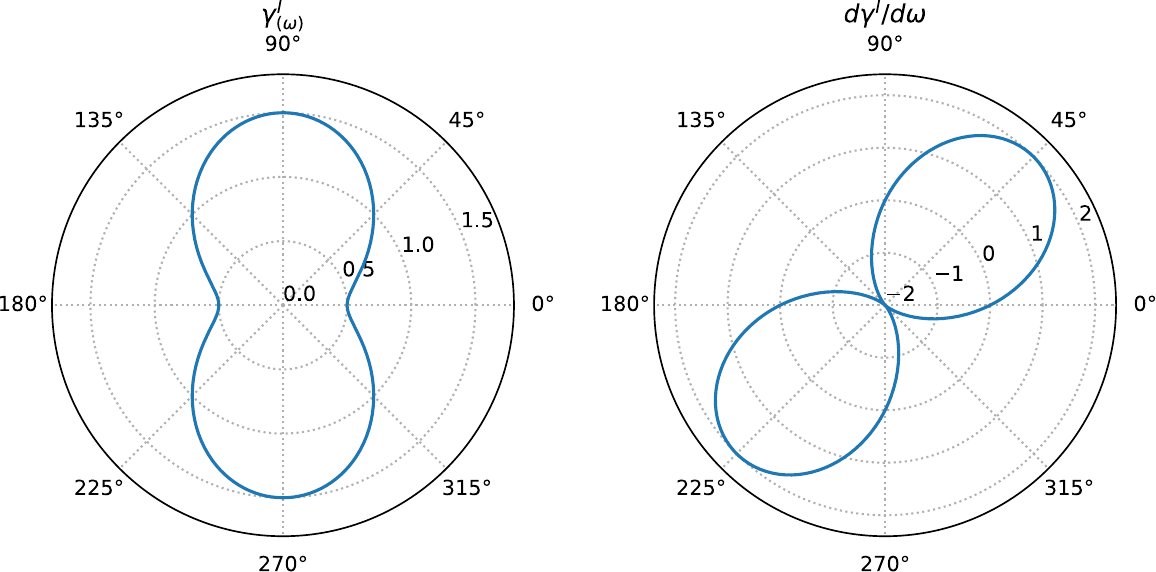}
\caption{Plot of $\gamma^{I}$ and $d\gamma^{I} / d\omega$ as a function of the inclination angle $\omega$ in polar coordinates.}
\label{FigPolarGamma}
\end{figure}

Grain boundary properties have been computed using material properties approximated to those of pure nickel at 1400 K ($\overline{\gamma} = 1~J \cdot m^{-2}$ and $\mu = 0.1~mm^{4} \cdot J^{-1}\cdot s^{-1}$). A Read-Shockley function \cite{Read1950} was used for the computation of $\gamma^{D}_{(\theta)}$ while the value of $\gamma^{I}_{(\omega)}$ was obtained thanks to the function $\gamma^{I}_{(\omega)}=0.5\cdot sin^2(\omega) + 1.5 \cdot cos^2(\omega)$ which maintains an average value of $\overline{\gamma}^{I}_{(\omega)}=1.0~J \cdot m^{-2}$. Figure \ref{FigPolarGamma} plots the value of $\gamma^{I}$ and $d\gamma^{I} / d\omega$ in polar coordinates, this function has no particular physical meaning, and it is used here to illustrate the impact of torque terms by considering reasonable variations of $\gamma^{I}_{(\omega)}$. \\

All simulations used an isotropic mesh size value of $h^{TRM}=4\cdot 10^{-4} mm$ (see Fig.\ref{Figure_Example_InitialTess} (right)) and a constant time step of $dt=5 s$. Three hours of thermal annealing were simulated. Compared to previous works \cite{Florez2020b, Florez2020c, Florez2020d}, these mesh size and time step values are smaller; this is because a higher precision on the description of grain boundaries and their dynamics is required when accounting for torque terms. Simulations without torque do not require such a level of detail, but the same numerical parameters were used in order to avoid precision bias while comparing the results.\\

\begin{figure}[!h]
\centering
\includegraphics[width=1.0\textwidth] {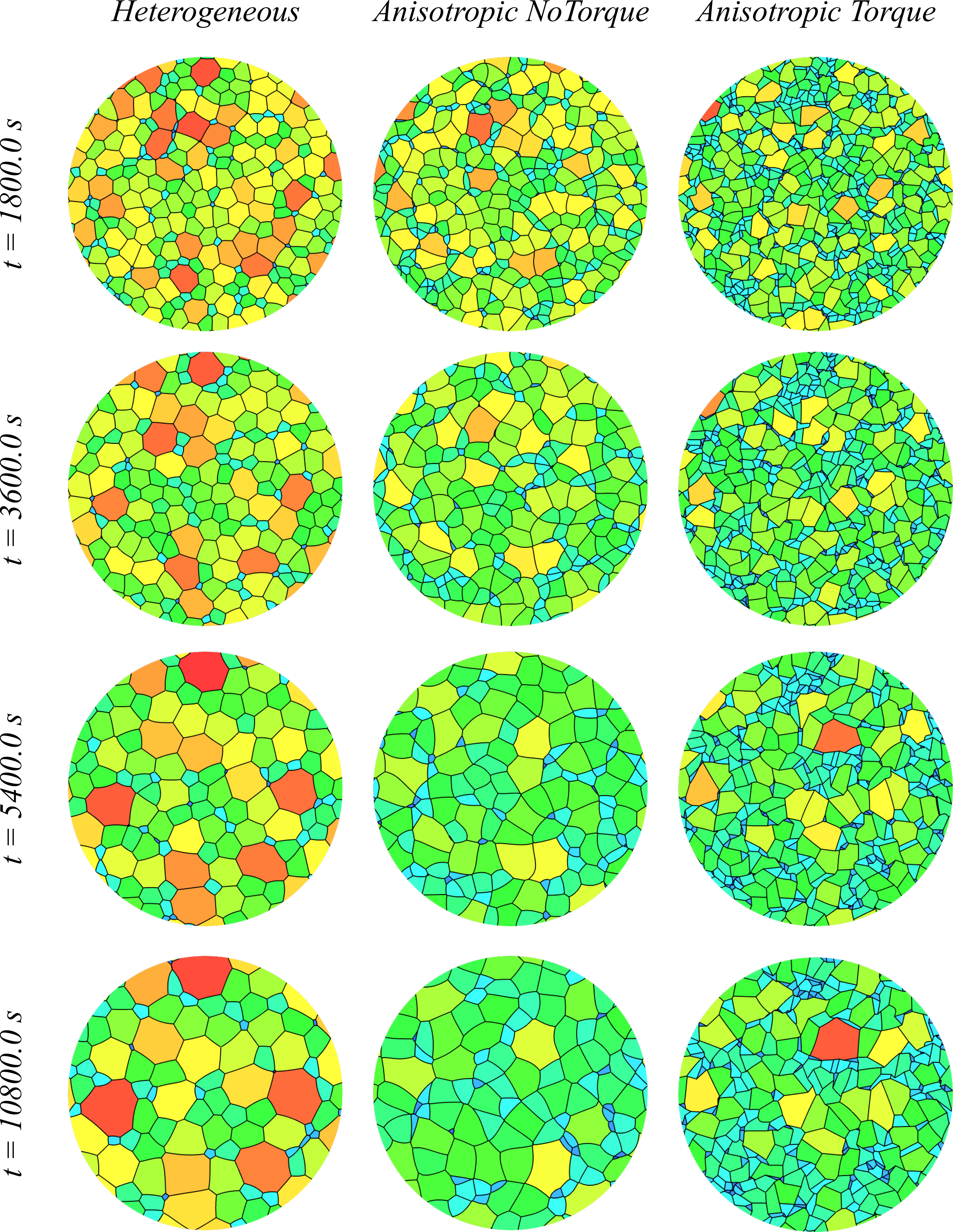}
\caption{Example (zoom) of the evolution of morphologies for the different tests as a function of time for the same initial microstructure. Very different morphologies are found for the tests, showing how influential torque terms can be. Colors are representative of the individual grain radius.}
\label{Figure_EvolutionTopology}
\end{figure}

\begin{figure}[!h]
\centering
\includegraphics[width=1.0\textwidth] {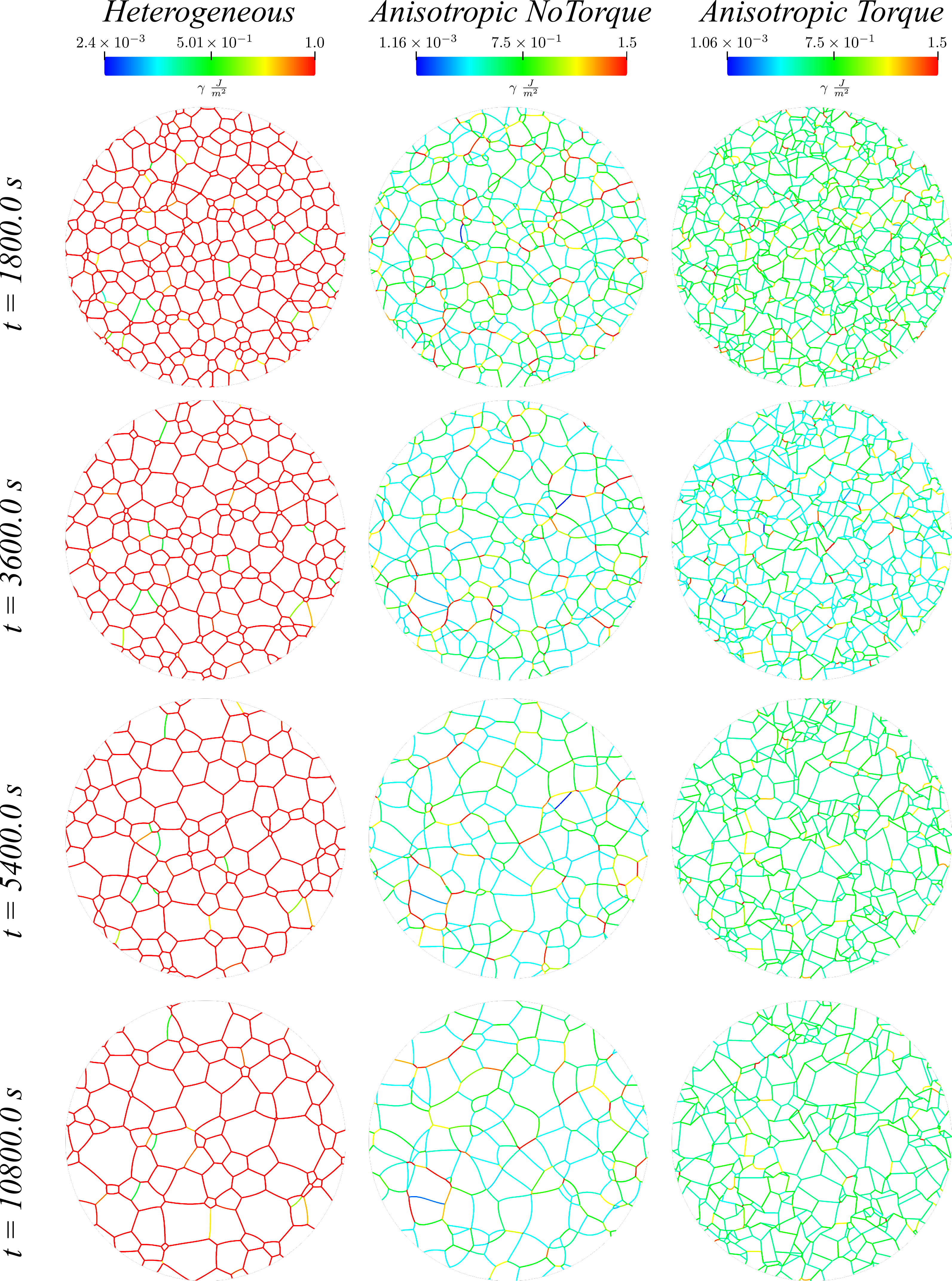}
\caption{Example (zoom) of the distribution of grain boundary energies in the domain for the different tests as a function of time for the same initial microstructure. Colors correspond to the GB energy averaged on each GB.}
\label{Figure_EvolutionTopology_2}
\end{figure}

Figures \ref{Figure_EvolutionTopology} and \ref{Figure_EvolutionTopology_2} illustrate one example (zoomed) of the microstructure evolutions for the three formulations by considering the same initial state. It is clear that the microstructure morphology is strongly related to the dependence of $\gamma$ to the inclination angle $\omega$ and to the use of torque terms into the computation of GB kinetics. The heterogeneous case (using a Read-Shockley formulation) behaves smoothly in all directions, driven by the Young's equilibrium of multiple junctions and conserving an equiaxed microstructure. The anisotropic case without torque tends to produce non-convex grain morphologies with smoothly curved grain boundaries. The evolution is driven not only by the Young's equilibrium of multiple junctions but also by the grain boundary energy minimization. Finally, the anisotropic case with torque terms, which remains driven by the energy minimization, behaves, however, in a very different way: 
equilibrium states at multiple junctions are highly affected by torque terms, as some junctions appear to have found their equilibrium at angles higher than $180^o$, this is not only given by the use of torque terms but also because of the $\gamma^{I}_{(\omega)}$ profile at these positions.\\

\begin{figure}[!h]
\centering
\includegraphics[width=1.0\textwidth] {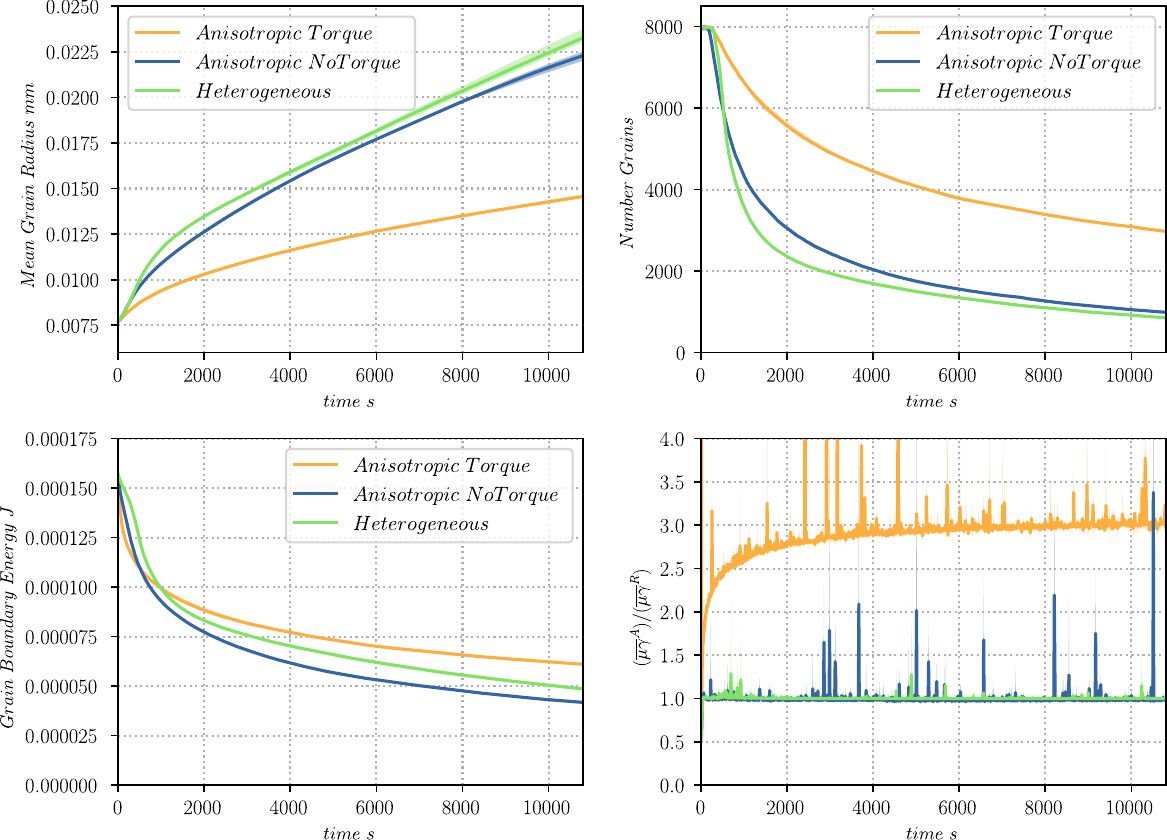}
\caption{Evolution of different averaged or integrated quantities as a function of time, (top-left) arithmetic mean grain radius , (top-right) number of grains, (bottom-left) total grain boundary energy and (bottom-right) ratio between the mean apparent reduced mobility and the real one (see Eq.\ref{Eq:Equation_Mean_Mobility}). Results for each case are averaged from the results of the different tessellations, the range of the results is also plotted on each figure with an alpha component of the same color.}
\label{Figure_Mean_Values}
\end{figure}

\begin{figure}[!h]
\centering
\includegraphics[width=0.7\textwidth] {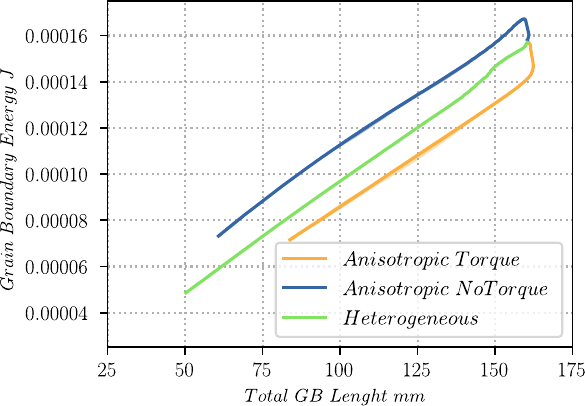}
\caption{Evolution of total GB energy as a function of the total GB length. Results for each case are averaged from the results of the different tessellations, the range of the results is also plotted on each figure with an alpha component of the same color.}
\label{Figure_Energy_Length}
\end{figure}

Mean values are also strongly affected by the use of the torque terms. Fig. \ref{Figure_Mean_Values}(top-left) illustrates the evolution of the arithmetic mean grain radius, which exhibits a much slower evolution for the anisotropic case considering the torque term than the other cases. Moreover, the evolution of the total grain boundary energy (Fig. \ref{Figure_Mean_Values}(bottom-left)) for the simulations not using torque terms shows a slightly higher total energy at the beginning of the simulation than simulations with torque. This tendency is inverted after 1000s of annealing, where the total energy of simulations with torque have a slower decrease rate, being at the end of the simulation, higher than their counterpart without torque. Of course, this does not mean that simulations with torque are less efficient when minimizing the GB energy. Figure \ref{Figure_Energy_Length} shows the total GB energy as a function of the total GB length, this plot illustrates how for the same length of boundaries, simulations with torque corresponds to the most efficient formulation in terms of overall energy minimization. This suggests that the fact that simulations with torque have higher total energies might be given by the path that boundaries take to obtain local minimums of boundary energy, which could slow down the GB network in an apparent higher energetic state than in the other formulations. Finally, the \emph{mean apparent reduced mobility} of the GBs, was computed considering only curvature flow:\\

\begin{equation}
\label{Eq:Equation_Mean_Mobility}
\centering
\overline{\mu\gamma}^A = \dfrac{\sum_i \int{ (\mu_i\gamma_i)^A  dl_i}}{\sum_i L_i} ,
\end{equation}

where $L_i$ is the total length of GB $i$ and the product $\mu_i\gamma_i$ is the \emph{apparent} reduced mobility of a given point $l_i$ on GB $i$:

\begin{equation}
\label{Equation_apparent_mobility}
\centering
(\mu_i\gamma_i)^A =\frac{|\vec{v_i}|} {\kappa_i},
\end{equation}

where the terms $|\vec{v_i}|$ and $\kappa_i$ are the velocity and mean curvature of GB $i$ evaluated at $l_i$ following the procedure described in \cite{Florez2020b}. Note that, one can also compute the \emph{real} value of the reduced mobility given by the product of the mobility $\mu_i$ and the local GB energy $\gamma_i$ at $l_i$, we will use the syntax $(\mu\gamma)^R$ to differentiate the \emph{real} from the \emph{apparent} reduced mobility.\\

Figure \ref{Figure_Mean_Values}(bottom-right) illustrates the form of the relationship between $\overline{\mu\gamma}^A$ and $\overline{\mu\gamma}^R$ during the simulation. Note how the heterogeneous and anisotropic tests without torque show values of around $(\overline{\mu\gamma}^A)/(\overline{\mu\gamma}^R) = 1.0$, with some sudden variations occurring when some grains disappear (where the computation of Eq.\ref{Equation_apparent_mobility} can deviate given the high values of curvature on small grains). Anisotropic simulations with torque show a more complex behaviour with values starting at $1.0$ and an equilibrium state at around $(\overline{\mu\gamma}^A)/(\overline{\mu\gamma}^R) = 3.0$. \\

\begin{figure}[!h]
\centering
\includegraphics[width=1.0\textwidth] {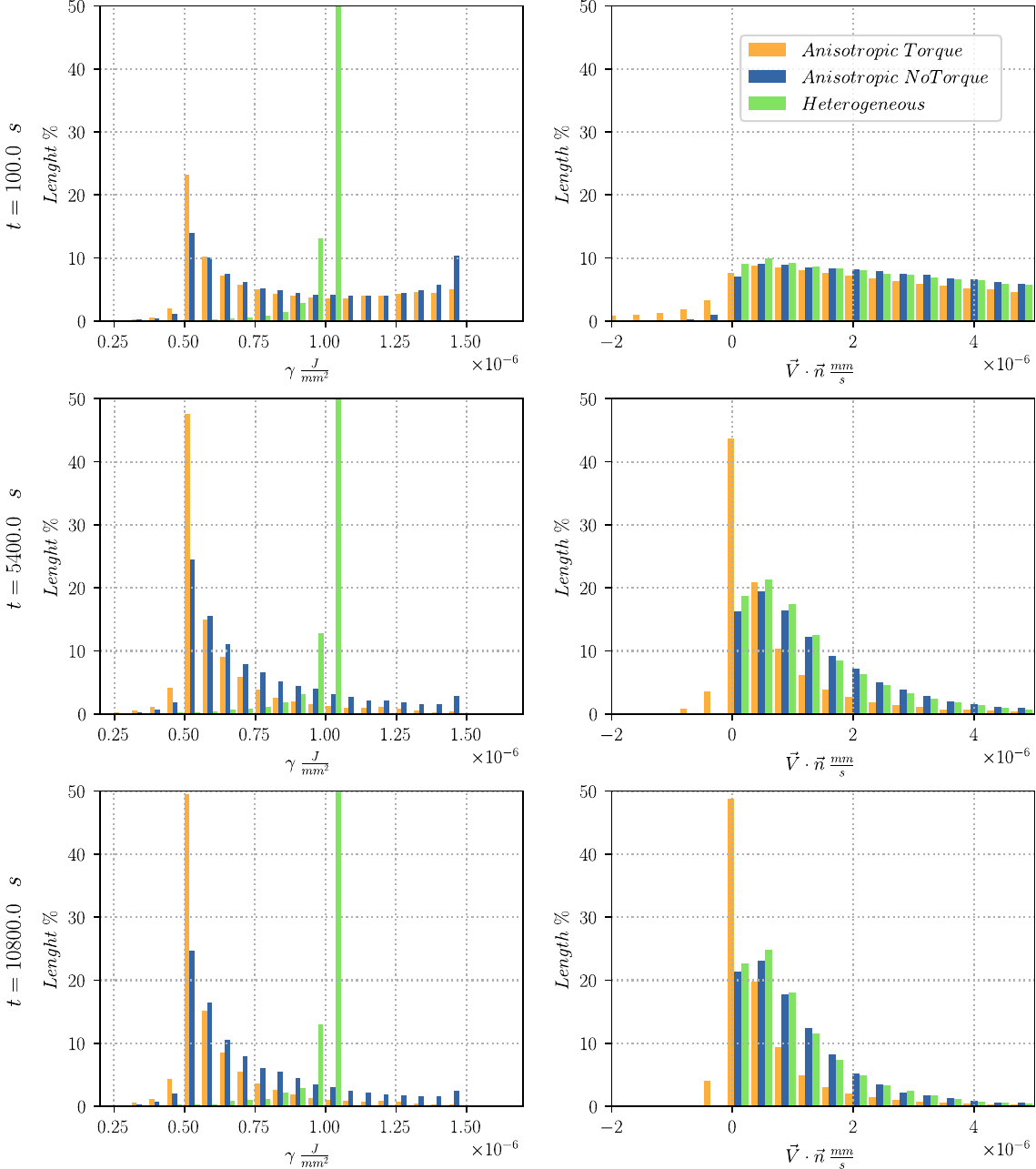}
\caption{Results of the test cases with different attributes on the computation of the velocity of grain boundaries. Each case uses four tests with four different initial tessellations. (left) Grain boundary energy distribution as a percent of the total length of GBs. (right) Scalar product between the velocity $\vec{v}$ and the local normal $\vec{n}$ as as a percent of the total length of GBs. }
\label{Figure_Join_Dot_Product}
\end{figure}

Fig.~\ref{Figure_Join_Dot_Product}(left) gives a deeper insight into the minimization of energy given by torque terms. Both anisotropic tests seem to promote low energy states by reducing the amount of interface with high energies. However, simulations with torque do so much faster, relatively to the total length of interfaces. (see also Figs.~\ref{Figure_EvolutionTopology} and \ref{Figure_EvolutionTopology_2}).\\

\begin{figure}[!h]
\centering
\includegraphics[width=1.0\textwidth] {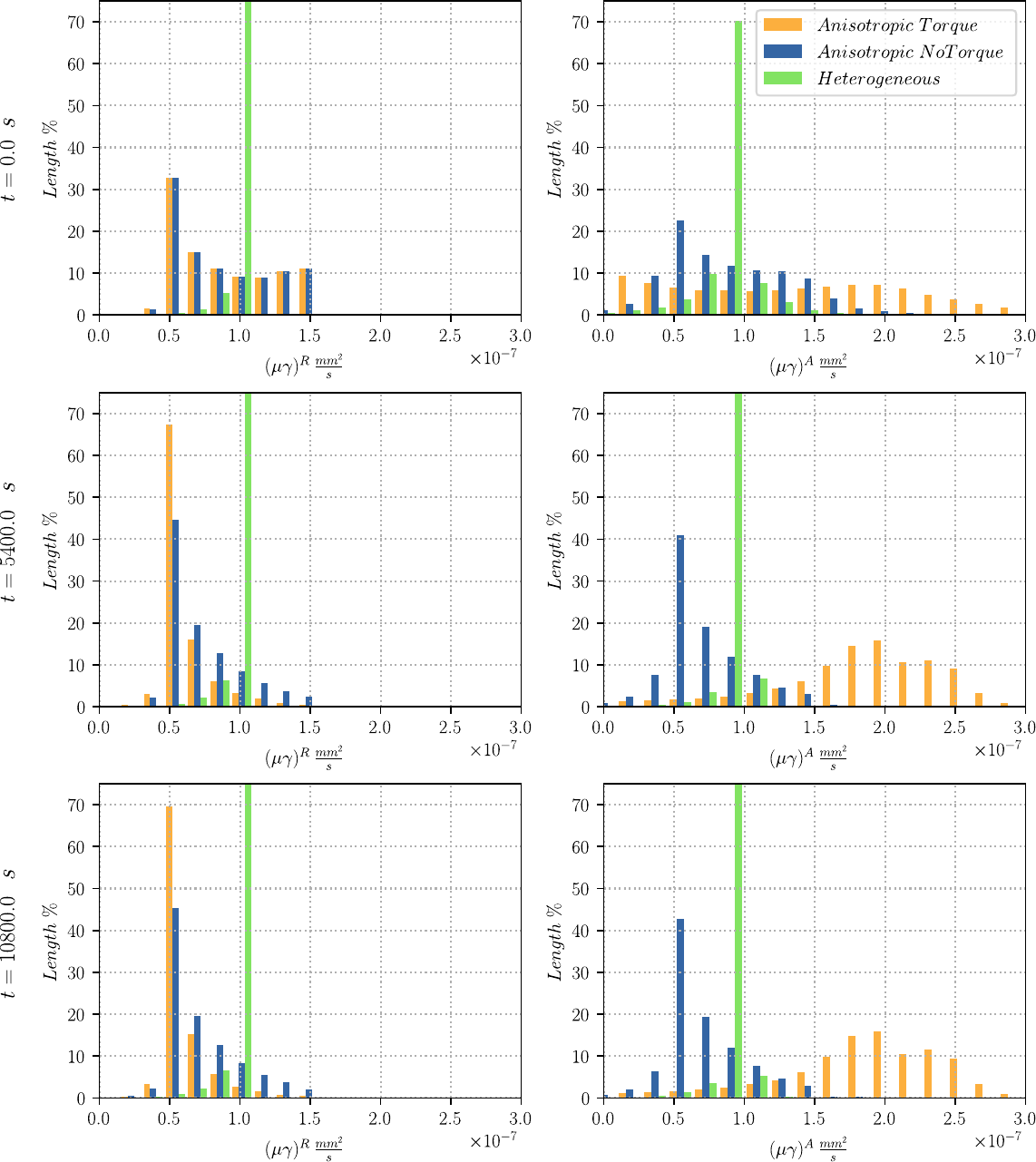}
\caption{Results of the test cases with different attributes on the computation of the velocity of grain boundaries. Each case uses four tests with four different initial tessellations. (left) \emph{Real} and (Right) \emph{apparent} reduced mobility distribution as a percent of the total length of GBs.}
\label{Figure_Join_Energy_Mean_Mobility}
\end{figure}

The behavior of $\overline{\mu\gamma}^A$ for the simulations with torque terms is explained by the fact that Eq.~\ref{Equation_apparent_mobility} underestimates the reduced mobility of stationary-curved GBs, while overestimates it on evolving-flat GBs, both mechanisms only appearing on simulations with torque. This is also observed in Fig.~\ref{Figure_Join_Energy_Mean_Mobility}(right) where the value $(\mu\gamma)^A$ is plotted in a distribution weighted by GB length at different times. GBs without torque present a values of $(\mu\gamma)^A$ lower than  around $1.5\times 10^{-7}~mm^2 \cdot s^{-1}$, while for GBs with torque, this value is distributed in a wider range with a maximun at around $2.0\times 10^{-7}~mm^2 \cdot s^{-1}$. Additionally, it is shown how values higher than $2.5\times 10^{-7}~mm^2 \cdot s^{-1}$ appear for these simulations, which are found for flat GBs with a velocity component higher than a curvature-flow type of velocity would suggest. Figure Fig.~\ref{Figure_Join_Energy_Mean_Mobility}(left) gives the distribution of $(\mu\gamma)^R$ for reference, showing how the only case that deviates on the computation of the apparent reduced mobility is the case with torque effects.
In some cases, torque effects driving the GBs in a direction contrary to the center of curvature are dominant, and thus produces velocities opposed to curvature-flow, this is illustrated in Fig.~\ref{Figure_Join_Dot_Product}(right) where the scalar product $\vec{v}\cdot\vec{n}$ finds negative values ($\vec{n}$ always pointing in the direction of the curvature center). This, however, only applies for simulations with torque effects. \\

Other statistical curves can be observed in Fig. \ref{Figure_Mackenzie_Distribs_MeanSize}. The disorientation distributions seem to maintain their \emph{Mackenzie} shape with a few deviations between 0° and 20°. These deviations are obtained thanks to the low GB energy for these disorientation values (Read-Shockley formulation), although this result is less predominant when using torque terms. Finally, the grain size distribution shows a clear difference between torque and no-torque simulations. Grain size evolves much quicker when torque effects are not used, which is a direct consequence of the high GB energy states, which disappear more slowly in no-torque simulations comparatively to simulations with torque terms (see figure \ref{Figure_Join_Energy_Mean_Mobility}(left)).

\begin{figure}[!h]
\centering
\includegraphics[width=1.0\textwidth] {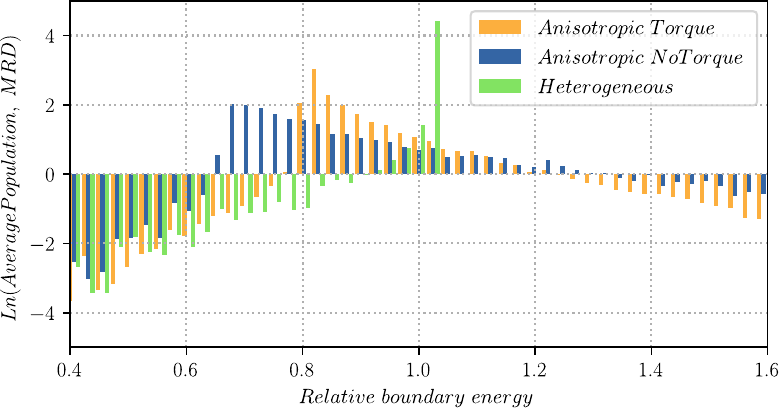}
\caption{Final relative boundary energy distribution in \emph{multiples of random density (MRD)} units. Results of the configuration using torque terms seem have the same tendency as the results reported in \cite{Beladi2013}. Energy data was normalized so the average relative value equals 1. }
\label{GregInverseCorrelationGamma}
\end{figure}

Finally, in \cite{Beladi2013} it has been reported \emph{a very strong inverse correlation between the grain boundary population and the relative grain boundary energy}. Here we have found a similar tendency for the anisotropic configurations (see Fig. \ref{GregInverseCorrelationGamma}), where the configurations with torque present a more pronounced slope. Also, the heterogeneous configurations showed a proportional behaviour (contrary to the observations in \cite{Beladi2013}).

\section{Discussion, conclusion and perspectives}\label{sec:conclusions}

In this article, we have measured the apparent reduced mobility of grain boundaries when heterogeneous ($\gamma_{(\theta)}$) and anisotropic ($\gamma_{(\theta, \omega)}$) properties are used on grain growth simulations. The apparent reduced mobility, in curvature flow meaning, of each interface was computed using Eq.\ref{Equation_apparent_mobility} which underestimates the reduced mobility of stationary-curved interfaces while overestimates it on evolving-flat interfaces (both mechanisms are only present on anisotropic simulations where torque terms are taken into account). \\

Torque effects have shown an apparent increase in the global reduced mobility of grain boundaries. Moreover, considering torque terms has produced a more coherent response regarding the minimization of the grain boundary energy, with a quicker \emph{relative} reduction of GBs with high energies. Two mechanisms are correlated to this behaviour: i. the faster shrinking of flat grain boundaries with high surface energy, and ii. the torque-driven equilibrium state of grain boundaries (stationary-curved) avoids high energy states produced by the inclination angle $\omega$.

While based on an empirical description of the inclination dependence, these results illustrate two main points highlighted in the introduction:
\begin{itemize}
    \item The apparent global and local reduced mobility described in Eqs. \ref{Eq:Equation_Mean_Mobility} and \ref{Equation_apparent_mobility} are classically discussed when reverse engineering of 3D in-situ experimental data is considered. As illustrated, the deviation from expected values and behaviour can potentially be explained by the fact that the GB stiffness must be taken into account in the analysis and not only the GB energy. This result may have the effect of weighting the increasingly common conclusion that this deviation is only due to the fact that reduced mobility cannot be properly described in the usual 5-dimensional space of misorientation and inclination. Of course, it is not proven here that this is not the case but it is clearly demonstrated that the torque terms must be taken into account in reverse engineering analysis of experimental data at the polycrystal scale.
    \item Another conclusion concerns the fact that GG full-field simulations, defined as anisotropic by considering the GB energy described in the usual 5-dimensional space, can potentially be misleading if the torque terms are not explicitly incorporated into the considered driving pressure.
\end{itemize}

Perspectives of these works concern the extension to 3D-polycrystals and the consideration of a more physical description of the inclination dependence and of the misorientation axis dependence. \\

\section*{Acknowledgments}
The authors gratefully acknowledge  ArcelorMittal, ASCOMETAL, AUBERT \& DUVAL, CEA,  FRAMATOME, SAFRAN, TIMET, Constellium and TRANSVALOR companies for their support through the DIGIMU consortium. 


\bibliography{ms}

\end{document}